\documentclass{emulateapj}

\usepackage{times}
\usepackage{color}
\usepackage{bm}
\usepackage{xspace}			
\usepackage[normalem]{ulem}
\usepackage{hyperref}

\newcommand{\beq}{\begin{equation}}
\newcommand{\eeq}{\end{equation}}
\newcommand{\beqn}{\begin{eqnarray}}
\newcommand{\eeqn}{\end{eqnarray}}
\newcommand{\beqno}{\begin{equation*}}
\newcommand{\eeqno}{\end{equation*}}
\newcommand{\beqnno}{\begin{eqnarray*}}
\newcommand{\eeqnno}{\end{eqnarray*}}

\newcommand{\GRay}{\texttt{GRay}\xspace}
\newcommand{\Odyssey}{\texttt{Odyssey}\xspace}	
\newcommand{\OdysseyEdu}{\texttt{Odyssey\underline{ }Edu}\xspace}

\newcommand{\Grids}{\textit{Grid}s\xspace}	
\newcommand{\Grid}{\textit{Grid}\xspace}	
	
\newcommand{\FPP}{\textit{frozen photon plane}\xspace}

\shorttitle{\Odyssey: A Public GPU-based GRRT Code in Kerr spacetime}

\begin{document}
\title{\Odyssey: A Public GPU-based Code for General-Relativistic Radiative Transfer in Kerr spacetime}
\author{Hung-Yi Pu \altaffilmark{1, 2}, Kiyun Yun \altaffilmark{1,3}, Ziri Younsi \altaffilmark{4,5}, and Suk-Jin Yoon\altaffilmark{3}}	

\affil{\altaffilmark{1}
	Equal first authors}
	
\affil{\altaffilmark{2}
	Institute of Astronomy \& Astrophysics, Academia Sinica, 11F of Astronomy-Mathematics Building, AS/NTU No. 1, Taipei 10617, Taiwan}

\affil{\altaffilmark{3}
	Department of Astronomy \& Center for Galaxy Evolution Research, Yonsei University, Seoul 120-749, Republic of Korea}			

\affil{\altaffilmark{4}
	Institut f\"ur Theoretische Physik, Max-von-Laue-Stra{\ss}e 1, D-60438 Frankfurt am Main, Germany}		
	
\affil{\altaffilmark{5}
	Mullard Space Science Laboratory, University College London, Holmbury St. Mary, Dorking, Surrey, RH5 6NT, UK}

\begin{abstract}
General-relativistic radiative transfer (GRRT) calculations coupled with the calculation of geodesics in the Kerr spacetime are an essential tool for determining the images, spectra and light curves from matter in the vicinity of black holes. Such studies are especially important for ongoing and upcoming millimeter/sub-millimeter (mm/sub-mm) Very Long Baseline Interferometry (VLBI) observations of the supermassive black holes at the centres of Sgr A$^{*}$ and M87.
To this end we introduce \Odyssey, a Graphics Processing Unit (GPU)-based code for ray tracing and radiative transfer in the Kerr spacetime. On a single GPU, the performance of \Odyssey can exceed 1 nanosecond per photon, per Runge-Kutta integration step.
\Odyssey is publicly available, fast, accurate, and flexible enough to be modified to suit the specific needs of new users. Along with a Graphical User Interface (GUI) powered by a video-accelerated display architecture, we also present an educational software tool, \OdysseyEdu, for showing in real time how null geodesics around a Kerr black hole vary as a function of black hole spin and angle of incidence onto the black hole.
\end{abstract}

\section{Introduction}
Theoretical studies of observable features in the strong gravity environment around black holes, such as emission line profiles \cite[e.g.,][]{mul04,fue04, fue07, you12}, reverberation \citep[e.g.,][]{rey99}, light curves of a hot spot \cite[e.g.,][]{sch04,bro05}, quasi-periodic oscillations (QPOs) \cite[e.g.,][]{sch04, sch06, fuk08}, and black hole shadow images \cite[e.g.][]{fal00, tak04}, provide useful insights into understanding the nature of these systems.
Recent mm VLBI observations of Sgr A$^{*}$ and M87 by the Event Horizon Telescope (EHT) helped to place constraints on the core size to within a few Schwarzschild radii, and demonstrated the potential capability of the EHT to directly image the shadow of a black hole \citep{doe08,doe12}. In the near future, images of black hole shadows are expected to be observed by mm/sub-mm VLBI observations, e.g., the EHT, BlackHoleCam and Greenland Telescope (GLT) projects \citep[]{ino14}.

As the resulting image and spectrum from black hole accretion and jets can vary due to the dynamics of the accretion flow \citep[e.g., free-fall vs. Keplerian sphere of plasma;][]{fal00, bro06}, and is also strongly dependent on the distribution of (both thermal and non-thermal) electrons \citep[e.g.][]{cha15, aki15}, a systematic study of the observational predictions of many different models throughout physically relevant parameter space is essential. As such, the rapid, efficient and accurate computation of both geodesics and the solution of the radiative transfer equations along these geodesics in the Kerr spacetime is a very important topic. 

The integration of geodesics in the Kerr spacetime can be performed either by using the transfer function method \citep[][]{cun75, fabian00}, the elliptic function method \citep[e.g.,][]{rau94, ago97, dex09, yang13}, or by direct numerical integration of the geodesic equations of motion \citep[e.g.][]{fue04, lev08, psa12,you12}. In the absence of scattering, the integration of the radiation transfer equation can be performed by dividing each ray into a series of small steps. Whilst the elliptic function method, being semi-analytic in form, is efficient for the calculation of emission from axisymmetric and optically thick objects like a geometrically thin accretion disk, it does not fare so well for systems which do not possess the necessary symmetry, such as the highly-turbulent, non-symmetric and magnetised flows found in GRMHD simulations of accretion onto black holes. Direct numerical integration of the geodesic equations of motion by the Runge-Kutta method is more suitable for GRRT computations as it makes no assumptions of the underlying geometry or thermodynamics of the accretion flow or spacetime being considered. Additionally, because a complex change of variables is not needed, the direct integration method is more straightforward to implement numerically and its incorporation into more sophisticated and physically-realistic models is transparent.

The calculation of geodesics and radiative transfer along each ray (geodesic) can be efficiently boosted through parallel computation. In capitalising on the advantages offered by parallel programming, much attention has been paid in recent years to the GPU. 
Being somewhat analogous to the messaging passing interface (MPI) model comprised of multiple CPUs, the GPU model is built around the concept of multiple Streaming Multiprocessors (SMs) containing several hundred threads. These threads work as a CPU does, i.e. concurrently, with a Single-Instruction, Multiple-Thread (SIMT) model on a single graphics card. However, a graphics card may comprise of several thousand processors. In order to take advantage of the architecture of the GPU, one of the major graphics card manufacturers, NVIDIA, released the Compute Unified Device Architecture (CUDA) platform for General-Purpose computing on Graphics Processing Units (GPGPU). Given ray-tracing, in the absence of scattering, is a trivially parallelizable problem (each ray may be treated as independent from all other rays), no communication between threads is necessary during the calculation of each ray. This makes the GPU model particularly appealing.

In this article we present a public, GPU-based GRRT code, \Odyssey, based on the ray-tracing algorithm presented in \citet{fue04}, and radiative transfer formulation described in \citet{you12}, implemented in CUDA C/C++. The performance on a single NVIDIA GPU graphics card exceeds one nano second per Runge-Kutta step per geodesic, similar or slightly better than that reported in another GPU-based ray-tracing code \GRay \citep{cha13}, in which a different ray-tracing algorithm is adopted. One of the direct applications of \Odyssey is for studying the spectra and images from black holes at horizon scales. This is an important observational goal of current and future mm/sub-mm VLBI observations of the accretion flows onto supermassive black holes. 

The article is organized as follows. In Section 2 the ray-tracing and radiative transfer formulation are derived and the numerical solution of this formulation is discussed. In Section 3 we introduce and discuss the GPU scheme on which \Odyssey is based. In Section 4 we present the results of several different benchmarking calculations for \Odyssey, including the calculation of images, spectra and lightcurves necessary for astrophysical calculations and comparisons with future observations. In Section 5 we assess the performance of \Odyssey, presenting timing benchmarks and comparisons with other ray-tracing codes. Section 6 is devoted to the summary and discussion.

\section{Formulation}
In this article we adopt the natural unit convention $(c=G=1)$, wherein the gravitational radius of a black hole of mass $M$ is given by $r_{\mathrm{g}}=M$. For a rotating (Kerr) black hole the spacetime metric may be written in Boyer-Lindquist coordinates as:
\begin{eqnarray}  
 \mathrm{d}s^2 &=& -\biggl( 1- {{2Mr} \over \Sigma}\biggr)\mathrm{d}t^2 
         - {{4aMr \sin^2\theta} \over \Sigma}\mathrm{d}t \ \! \mathrm{d}\phi 
         + {\Sigma \over \Delta}\mathrm{d}r^2 \nonumber \\ 
     & & \hspace{-0.25cm} + \Sigma \ \!  d\theta^2  
    + \biggl( r^2+a^2 + {{2a^2Mr \sin^2\theta} \over \Sigma} \biggr) 
                  \sin^2\theta \ \! \mathrm{d}\phi^2 \ , 
\end{eqnarray}  
  where $\Sigma \equiv r^2+a^2\cos^2\theta$ and $\Delta \equiv r^2 -2Mr +a^2$. Hereafter the black hole mass is set to unity, which is equivalent to normalising the length scale to $r_{\mathrm{g}}$ and the timescale to $r_{\mathrm{g}}/c$.
From the separability of the Hamilton-Jacobi equations for a Kerr black hole, the geodesic equations of motion may be reduced to a problem of quadratures (four constants of motion (see Section 2.1) and four ODEs) with the variables ($\dot{t}$, $\dot{r}^{2}$, $\dot{\theta}^{2}$, $\dot{\phi}$), where an overdot denotes differentiation with respect to the affine parameter.
Complications related to the uncertainty in the signs of $r$ and $\theta$ at turning points in the geodesic motion arise from this approach, and so to circumvent this issue we instead consider the second derivatives of $r$ and $\theta$, and integrate six differential equations for $(\dot{r}, \dot{\theta}, \dot{\phi}, \dot{t}, \dot{p_{r}}, \dot{p_{\theta}})$, where $(p_{r}, p_{\theta})$ are components of the covariant four-momentum of the geodesic \citep{fue04}. 

Such an algorithm is sufficiently straightforward to implement into GRRT calculations. The covariant four-momentum $p_{\alpha}$ is computed at each integration step, and can be directly used to compute the relative energy shift, $\gamma$, between radiation emitted from material circulating around the black hole with four-velocity $u^{\alpha}$ and radiation received by a distant observer:
\begin{equation}
\label{eqn:redshift}
\gamma\equiv\frac{\nu}{\nu_0}=\frac{p_{\alpha}u^{\alpha}|_{\infty}}{p_{\alpha}u^{\alpha}|_{\lambda}} \;.
\end{equation}
The scalar product $p_{\alpha}u^{\alpha}$ is a frame-invariant quantity which may be calculated in any desired frame at that position in spacetime. For simplicity, we choose to evaluate $p_{\alpha}u^{\alpha}$ in the observer reference frame at that particular point along the ray (indicated by the affine parameter, $\lambda$). Subscripts ``$0$" and ``$\infty$" denote quantities evaluated in the local fluid rest frame and in the reference frame of a distant (stationary) observer, respectively. For completeness, in this section we summarise the governing differential equations \citep{fue04}, initial conditions, and radiative transfer formulation \citep{you12} used in \Odyssey.

\subsection{Ray-Tracing Algorithm}

The Kerr spacetime is of Petrov-type D and, being independent of both $t$ and $\phi$ coordinates, possesses two Killing vectors. These give rise to the conservation of energy, $E$, and conservation of angular momentum, $L_{\mathrm{z}}$, where $L_{\mathrm{z}}$ is the projection of the particle angular momentum along the black hole spin axis. The rest mass, $\mu$, of the particle (0 for photons and massless particles and -1 for particles with mass) and the Carter constant, $Q$, are also conserved along each geodesic. As such, the Kerr black hole possesses four constants of motion. From the Lagrangian, the covariant four-momenta components of a geodesic may be written as:
\begin{eqnarray}
  p_{\rm t} &=& -E\ ,\\
\label{pr}
  p_{\rm r} &=& \frac{\Sigma}{\Delta} \dot r\ , \\
\label{ptheta}
  p_{\rm \theta}  &=& \Sigma \dot \theta\ , \\
  p_{\rm \phi} &=& L_{\mathrm{z}}\;.
\end{eqnarray}
In addition,
$E$ and $L_{\mathrm{z}}$ may be derived from the initial conditions of the ray via the following formulae:
 \begin{eqnarray}
  E^{2} &=& \left(\frac{\Sigma-2r}{\Sigma\Delta}\right) \left(\Sigma\dot{r}^{2}+\Sigma\Delta\dot{\theta}^{2} - \Delta \mu \right)+\Delta\dot{\phi}^{2}\sin^{2}\theta \; , \\
 L_{\mathrm{z}} &=& \frac{(\Sigma\Delta\dot{\phi}-2arE)\sin^{2}\theta}{\Sigma-2r}\;.
 \end{eqnarray} 
The corresponding geodesic equations of motion may then be written as:
\begin{eqnarray}
\label{tdot}
  \dot t & = & E + \frac{2 r(r^2+a^2)E-2ar \ \! L_{\mathrm{z}}}{\Sigma\Delta}\ ,\\  
\label{rdot2}
  \dot r^2 & = & {\Delta \over \Sigma} 
    \big(\mu + E \dot t - L_{\mathrm{z}} \dot \phi- \Sigma \dot \theta^2 \big)\ , \\ 
\label{thetadot2}
  \dot \theta^2 & = & {1 \over \Sigma^2} \big[ 
      Q + (E^2 + \mu) a^2  \cos^2\theta - L_{\mathrm{z}}^2 \cot^2 \theta \big]\ ,  \\ 
\label{phidot}  
  \dot \phi & = & {{2a rE  + (\Sigma - 2 r)L_{\mathrm{z}}\csc^2\theta} \over 
    {\Sigma\Delta} } \ ,   
\end{eqnarray}    
  where
\begin{equation}
  Q \equiv p_{\rm \theta}^{2}+\left[L_{\mathrm{z}}^{2}\csc^{2}\theta-a^{2} (E^{2}+\mu) \right]\cos^{2}\theta\;.
\end{equation}
Equations (\ref{rdot2}) and (\ref{thetadot2}) may be replaced with two equations for the covariant four-momenta \citep[for details, see][]{fue04} as follows:
\begin{eqnarray}
\label{rdot}
   \dot p_{\rm r} &=& 
   \frac{1}{\Sigma \Delta} \Big\{ \left[\left(r^{2}+a^{2} \right)\mu - \kappa \right](r-1) + r\Delta\mu \nonumber \\
  & &+ 2r(r^2+a^2)E^2-2aEL_{\mathrm{z}} \Big\} -\frac{2{p_r}^2(r-1)}{\Sigma}\ ,\\
\label{thetadot}
  \dot p_{\rm \theta} &=& \frac{\sin\theta\cos\theta}{\Sigma}
   \left[\frac{L^2}{\sin^4\theta}-a^2\left(E^2 + \mu\right)\right]\ ,
\end{eqnarray}
  where $\kappa \equiv Q+L_{\mathrm{z}}^2+a^2(E^2 + \mu)=p_{\rm \theta}^{2}+L_{\mathrm{z}}^{2}\csc^{2}\theta +a^{2}\left(E^{2}\sin^{2}\theta + \mu \right)$.
  The final equations of motion for the six parameters $(r, \theta, \phi, t, p_{r}, p_{\theta})$ are
then given by equations (\ref{pr}), (\ref{ptheta}), (\ref{tdot}), (\ref{phidot}), 
  (\ref{rdot}) and (\ref{thetadot}). 
 
In \Odyssey, a fifth-order Runge-Kutta scheme with adaptive step size \cite[][]{pre92} is used to integrate these equations. In addition, to avoid numerical problems when a photon passes too close to the pole $(\theta=0,\pi)$, $\sin\theta$ is set to be $10^{-8}$ when $\sin\theta<10^{-8}$. As shown later in \S4, such a consideration is acceptable in practical ray-tracing and GRRT calculations. 
\subsection{Initial Conditions}
For an observer who receives the ray at an inclination angle $\theta_{\mathrm{obs}}$ and radial position $r_{\mathrm{obs}}=\infty$, 
the celestial coordinates in the observer's image frame, $(\alpha,\beta)$, are calculated as \citep{cha83}:
\begin{equation}
\alpha=L_{\mathrm{z}}\csc\theta_{\mathrm{obs}}\;,
\end{equation}
\begin{equation}
\beta=\sqrt{Q/E^{2}+a^2\cos^{2}\theta_{\mathrm{obs}}-L_{\mathrm{z}}^{2}\cot^{2}\theta_{\mathrm{obs}}}\;,
\end{equation}
where $\beta$ is also equal to the initial value of $-p_{\theta}$.
However, we wish to define an observer at some arbitrary position $(r,~\theta,~\phi)$ in space
(not necessarily infinitely removed from the black hole). Such an approach has several advantages,
including reducing the time needed to integrate the geodesic from the observer to the black hole.
The initial conditions of a ray arriving in this observer's image plane are calculated as follows.

The observer grid is constructed as a left-handed rectangular coordinate system with the 
$z$-axis oriented towards the black hole centre. The observer's axes are denoted by 
$\overline{\mathbf{x}}\equiv\left(x,\ y,\ z \right)^{\mathbf{T}}$. The black hole coordinate system is right-handed,
rectangular, and denoted by $\overline{\mathbf{x}}^{\prime}\equiv\left(x^{\prime},\ y^{\prime},\ z^{\prime} \right)^{\mathbf{T}}$.
The observer is located at a distance $r_{\mathrm{obs}}$ from the black hole centre, at an angle $\theta_{\mathrm{obs}}$
from the positive black hole $z^{\prime}$-axis (coinciding with the spin axis) and at an angle $\phi_{\mathrm{obs}}$ with
respect to the black hole's $x^{\prime}$-axis. Whilst the value of $\phi_{\mathrm{obs}}$ is arbitrary, since the Kerr
metric does not depend on $\phi$, there are situations where specifying the observer's azimuthal position is important. For
instance, in time-dependent radiation transfer calculations or imaging fully three-dimensional anisotropic accretion 
flows such as those found in state-of-the-art GRMHD simulations \cite[e.g.,][]{mck13}, one may wish to resolve
particular local and perhaps transient features, e.g. outflows, magnetic reconnection events and shocks.

It is assumed that the observer's image plane is a two-dimensional grid with zero curvature and that
all rays received by the observer arrive perpendicular to this grid. Close to the black hole, spacetime curvature becomes
significant and the observer image plane must possess some curvature-dependent distortion, which would in turn
distort the calculated image. Rays would no-longer arrive perpendicular to the image plane. To image a black hole closer to the event horizon would require defining an appropriate orthonormal tetrad basis in which to place the observer \citep[e.g.][]{bar72, mar96}. Since we concern ourselves only with calculating what a distant observer would actually observe, we place our observer at a distance of $10^{3}~r_{\mathrm{g}}$ from the black hole. At this distance the deviation of geodesics from Minkowski spacetime is smaller than the numerical precision used
to integrate the geodesic itself. Therefore, the spacetime may be taken as Euclidean and we can safely employ the reverse ray-tracing method
under the aforementioned assumptions.

In order to determine the initial conditions of rays starting on the observer's grid, the observer coordinate system 
$\overline{\mathbf{x}}$ must be transformed into the black hole coordinate system $\overline{\mathbf{x}}^{\prime}$.
This may be accomplished through the following series of transformations: (i) rotate clockwise by $\left(\pi-\theta_{\mathrm{obs}} \right)$ about the $x$-axis ($\mathrm{\mathbf{R}}_{x}$), (ii) rotate clockwise by $\left(2\pi-\phi_{\mathrm{obs}} \right)$ about
the $z$-axis ($\mathrm{\mathbf{R}}_{z}$), (iii) reflect in the plane $y=x$ ($\mathrm{\mathbf{A}}_{y=x}$), (iv) translate $\overline{\mathbf{x}}$ so that the origins of both coordinate
systems coincide ($\mathrm{\mathbf{T}}_{\mathrm{\mathbf{x}\rightarrow}\mathbf{x}'}$). This may be calculated as follows:

\begin{eqnarray}
\mathbf{x}' &=& \mathrm{\mathbf{A}}_{y=x} \ \! \mathrm{\mathbf{R}}_{z}(2\pi-\phi_{\mathrm{obs}})\mathrm{\mathbf{R}}_{x}(\pi-\theta_{\mathrm{obs}})\mathbf{x}+\mathrm{\mathbf{T}}_{\mathrm{\mathbf{x}\rightarrow}\mathbf{x}'} \nonumber \\
&=& \left(
\begin{array}{c}
\mathcal{D}(y,z)\cos\phi_{\mathrm{obs}}-x\sin\phi_{\mathrm{obs}}  \\
\mathcal{D}(y,z)\sin\phi_{\mathrm{obs}}+x\cos\phi_{\mathrm{obs}}  \\
 \left(r_{\mathrm{obs}}-z\right)\cos\theta_{\mathrm{obs}}+y\sin\theta_{\mathrm{obs}}  \\
\end{array}
\right) \ , \label{x_prime}
\end{eqnarray}
where
\begin{equation}
\mathcal{D}(y,z) \equiv \left(\sqrt{r_{\mathrm{obs}}^{2}+a^{2}}-z\right)\sin\theta_{\mathrm{obs}}-y \cos\theta_{\mathrm{obs}} \ .
\end{equation}
The transformation from Cartesian coordinates to Boyer-Lindquist coordinates is given by
\begin{eqnarray}
r &=& \sqrt{\frac{w+\sqrt{w^{2}+4a^{2}z'^{2}}}{2}} \ , \label{CartBLr} \\
\theta &=& \mathrm{arccos}\left(\frac{z'}{r} \right) \ , \label{CartBLtheta} \\
\phi &=& \mathrm{atan2}\left(y',x' \right) \ , \label{CartBLphi}
\end{eqnarray}
where $w\equiv x'^{2}+y'^{2}+z'^{2}-a^{2}$.
Substituting the components of equation (\ref{x_prime}) into equations (\ref{CartBLr})--(\ref{CartBLphi}) gives the initial
$(r,~\theta,~\phi)$ conditions for a photon on the observer grid. 

Next the initial velocities of the ray must be determined. Each ray arrives perpendicular to the image plane, 
moving parallel to the $z$-axis, hence we set $\left(\dot{x},~\dot{y},~\dot{z} \right)=(0,~0,~1)$. Subsequent differentiation of
equation (\ref{x_prime}) yields the Cartesian components of the ray's velocity in black hole coordinates:
\begin{equation}
\dot{x}' = \left(
\begin{array}{c}
-\sin\theta_{\mathrm{obs}}\cos\phi_{\mathrm{obs}} \\
-\sin\theta_{\mathrm{obs}}\sin\phi_{\mathrm{obs}} \\
-\cos\theta_{\mathrm{obs}} \\
\end{array}
\right) \ . \label{grid_velocity}
\end{equation}
Finally, to obtain the ray's velocity components in Boyer-Lindquist coordinates we differentiate equations 
(\ref{CartBLr})--(\ref{CartBLphi}) with respect to affine parameter, solve for $(\dot{r},~\dot{\theta},~\dot{\phi})$,
and substitute for equation (\ref{grid_velocity}). Upon simplification this yields:
\begin{eqnarray}
\dot{r} &=& -\frac{r \mathcal{R} \sin\theta \sin\theta_{\mathrm{obs}} \cos \Phi+\mathcal{R}^{2} \cos\theta \cos\theta_{\mathrm{obs}}}{\Sigma} \ , \label{rdot_initial} \\
\dot{\theta} &=& \frac{r \sin\theta \cos \theta_{\mathrm{obs}} - \mathcal{R} \cos\theta \sin\theta_{\mathrm{obs}} \cos \Phi}{\Sigma} \ , \label{thetadot_initial}\\
\dot{\phi} &=& \frac{ \sin\theta_{\mathrm{obs}} \sin \Phi}{\mathcal{R} \sin\theta} \ , \label{phidot_initial}
\end{eqnarray}
where $\mathcal{R}\equiv \sqrt{r^2+a^2}$ and $\Phi\equiv \left(\phi -\phi_{\mathrm{obs}}\right)$. We now have initial conditions for $(r,~\theta,~\phi,~t,~p_{r},~p_{\theta})$.

\subsection{Radiative Transfer}
To compute the jet emission and image in the observers' reference frame, one can use the ray-tracing scheme outlined in \S2.1 to trace the received ray backwards in time with initial conditions as described in \S2.2. 
The frequency shift is related to the plasma motion and varies from point to point along the ray, cf. equation (\ref{eqn:redshift}). Consequently, a frequency correction from the observed frequency to the local frequency is required at every point along the ray, because physical process take place in the local co-moving frame.

Along each ray the covariant GRRT
equation may be written as: 
\begin{equation}\label{eq:grrt0}
\frac{\mathrm{d}\mathcal{I}}{\mathrm{d}\tau_{\nu}}=-\mathcal{I}+\frac{\eta}{\chi}\;,
\end{equation}
where the Lorentz-invariant intensity ($\mathcal{I}$) is related to the specific intensity ($I$) as $\mathcal{I}=I_{\nu}/\nu^{3}=I_{\nu_0}/\nu_0^{3}$, where $\nu$ is the frequency of radiation.
The invariant absorption coefficient ($\chi$) and invariant emission coefficient ($\eta$) are given by
$\chi=\nu\alpha_{\nu}$ and $\eta=j_{\nu}/\nu^{2}$ respectively, where $\alpha_{\nu}$ and $j_{\nu}$ are respectively the specific emission and absorption coefficient evaluated at a frequency $\nu$. The optical depth of the medium at a given frequency $\nu$ is denoted by $\tau_{\nu}$.

Defining the source function $\mathcal{S}\equiv \eta/\chi$, equation (\ref{eq:grrt0}) may be directly integrated, yielding:
\begin{equation}\label{eq:sol}
\mathcal{I}(\tau_{\nu})=\mathcal{I}_{0}e^{-\tau_{\nu}}+\int^{\tau_{\nu}}_{\tau_{0}} \mathcal{S}(\tau'_{\nu}) e^{-(\tau_{\nu}-\tau'_{\nu})}d\tau'_{\nu}\;,
\end{equation}
where the optical depth $\tau_{\mathrm{\nu}}$ is given by
\begin{equation}\label{eq:tau}
\tau_{\nu}(\lambda)=-\int^{\lambda}_{\lambda_{0}} \alpha_{0,\nu}(\lambda')k_{\alpha}u^{\alpha}|_{\lambda'}d\lambda'\;,
\end{equation}
and $\lambda$ is the affine parameter.

By combining equations (\ref{eq:sol}) and (\ref{eq:tau}),
the solution of the radiative transfer equation can be can be reduced to two decoupled differential equations \citep{you12}
\begin{eqnarray}
\frac{\mathrm{d}\tau}{\mathrm{d}\lambda} &=& \gamma^{-1}\alpha_{0,\nu}\;, \label{eq:grrt1} \\
\frac{\mathrm{d}\mathcal{I}}{\mathrm{d}\lambda} &=& \gamma^{-1}\left(\frac{j_{0,\nu}}{{\nu}^{3}}\right)\;\exp(-\tau)\;, \label{eq:grrt2}
\end{eqnarray}
which are easily integrated along with the geodesic as the ray is propagated backwards in time.

\section{GPU scheme}
GPUs enable high multithreading and are designed for massively parallel computation\footnote{Physically, a GPU is built in an array of Streaming Multiprocessors (SMs). Each SM has N cores, or Streaming Processors (SP), and each SP is massively threaded. Typical CUDA-capable GPUs consist of several hundred to several thousand cores.}.
Using a mapping of one CUDA thread to one pixel, the method of parallel computation done by CUDA can be illustrated in Figure \ref{fig:gpuwork}.

In this analogy, the observer's image plane is composed of $N_{\alpha}\times N_{\beta}$ (hereafter assume $N_{\alpha}= N_{\beta}=N$) pixels in the $\alpha$ and $\beta$ directions, and is decomposed into multiple \Grids. 
Each \Grid is subdivided into a two-dimensional hierarchy of blocks and threads, and a CUDA kernel function run on each GPU is performed, \Grid by \Grid, until the entire image plane is covered.
To ensure there are enough threads to cover an image composed of $N^{2}$ pixels, the number of \Grids in each direction ($\alpha,\beta$) must satisfy the following condition:

\begin{eqnarray}
	\label{StencilDim}
		k_{i}= \left \lceil {{N} \over {Dim_{i}^{Block} \times Dim_{i}^{Grid}}} \right \rceil \ .
	\end{eqnarray}  
Note again that $Dim^{Grid}$ represents the number of blocks, $Dim^{Block}$ represents the number of threads and $i=(\alpha, 
\beta)$ is the dimension index. The notation $\lceil k \rceil$ is the smallest integer not less than $k$ (the `ceiling' function).

The global memory on each GPU is allocated by the CUDA command {\texttt{cudaMalloc()}, which is composed of two parts.
First, for each working thread, some global memory is required for saving specific variables until each thread finishes its computation. 
Let $P$ be the corresponding memory for each thread, then an amount $P \times T$ of memory is required, where $T$ is the number of working threads.
Second,
part of the global memory must be assigned for saving the computed data which will later be returned to the CPU.
For each pixel, if $Q$ is the amount of information for each pixel to be later transported from CUDA to the host CPU, then $Q \times N^{2}$ of memory is required for the entire image.
 By manipulating the value of $P$ and $Q$, the user can easily design their own task and assign multiple outputs (e.g. $\alpha$, $\beta$, $\gamma$, $I_{\nu}$, \ldots) to each pixel of the image for subsequent calculations.

\begin{figure} 
\begin{center}
\includegraphics[width=0.45\textwidth]{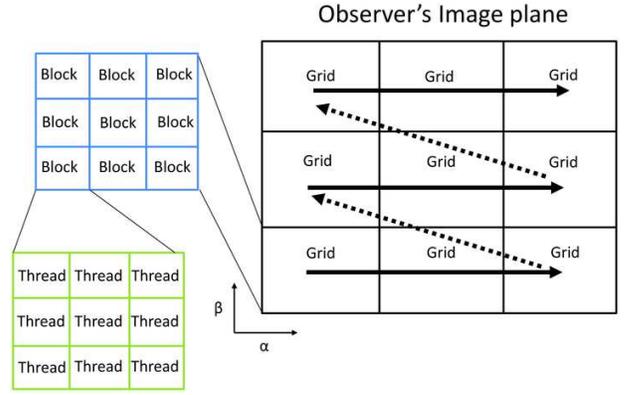}
\caption{Mapping the threads to pixels of image, as discussed in Section 3.} \label{fig:gpuwork}
\end{center}
\end{figure}

\section{Benchmark Astrophysical Calculations}

\subsection{Grid Projection}
In this section we perform several standard benchmark tests of the performance of \Odyssey. First, we calculate the projection of an evenly spaced rectangular grid in the observer's image plane $(\alpha,\beta)$ onto the equatorial plane of a Kerr black hole \citep[see][]{sch04, dex09, cha13} via the following coordinate transformation:
\begin{equation}\label{eq:x}
x=\sqrt{r^2+a^2}\cos\phi\;,
\end{equation}
\begin{equation}\label{eq:y}
y=\sqrt{r^2+a^2}\sin\phi\;.
\end{equation}
This is illustrated in Figure \ref{fig:test_projection}, which recovers the results of the aforementioned previous studies, e.g., Figure 2 of \citet{sch04} and Figure 3 of \citet{dex09}. Note that the horizontal and vertical axis in Figure \ref{fig:test_projection} corresponds to the $-\beta$ and $-\alpha$ direction, as adopted by 
those authors.

\begin{figure} 
\begin{center}
\includegraphics[width=0.5\textwidth]{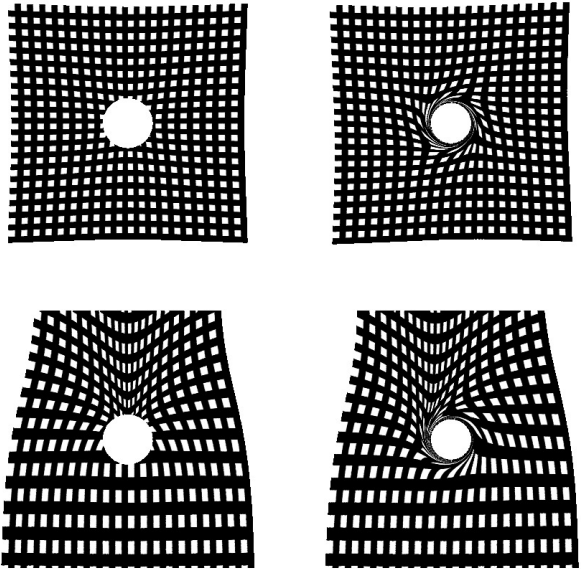}
\caption{Projection of uniform grids in the observer's frame $(\alpha,\beta)$ onto the equatorial plane $(x,y)$ of a non-rotating ($a=0$; left panels) and rapidly rotating ($a=0.95$; right panels) black hole. The observer inclination angle is set to be $i\simeq0^{\circ}$ (top panels) and $i=60^{\circ}$ (bottom panels). This figure recovers the results of previous studies, e.g., Figure 2 of \citet{sch04} and Figure 3 of \citet{dex09}. The horizontal and vertical axis  correspond to the $-\alpha$ and $-\beta$ direction, as adopted by the aforementioned authors.}\label{fig:test_projection}
\end{center}
\end{figure}

In Figure \ref{fig:test_projection2} we also consider a more illustrative example by projecting an evenly-spaced grid in the equatorial plane $(x,y)$ of the black hole onto the observer's image plane $(\alpha,\beta)$ for non-spinning ($a=0$, left) and spinning ($a=0.95$, right) black holes. The observer's image frame is shown in Figure \ref{fig:test_projection2}, therefore the horizontal and vertical axes simply correspond to the $\alpha$ and $\beta$ directions.  
The bending of rays emitted around the black hole result in an expansion of the grid on the equatorial plane, as can be clearly seen in the face-on case (top panels).  Compared with the grid in front of the black hole, the thickness of the grid behind is magnified by gravitational lensing (bottom panels). When the black hole is rotating (right panels), the rays are dragged by the rotation of the spacetime around the black hole (the frame-dragging effect). This effect results in the grid being distorted in an anti-clockwise direction (i.e. in the same direction as the rotation of the black hole).

\begin{figure} 
\begin{center}
\includegraphics[width=0.5\textwidth]{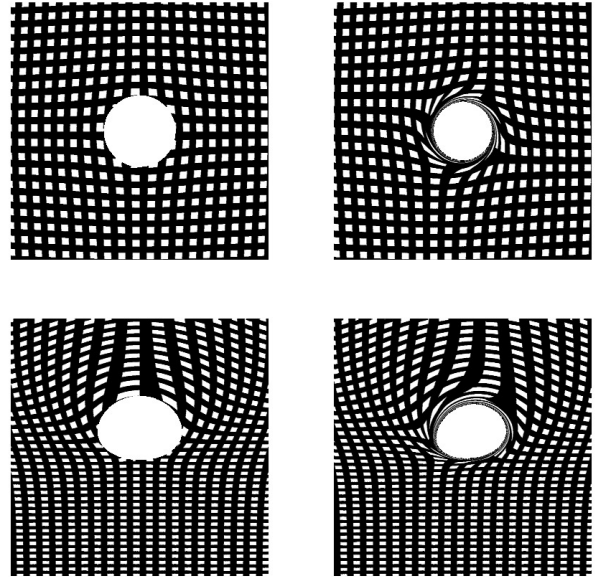}
\caption{Projection of uniform grids in the equatorial plane $(x,y)$ onto the observer's image plane $(\alpha,\beta)$, with the same parameters as used in Figure \ref{fig:test_projection}. The size of the field-of-view is $10~r_{g} \times 10~r_{g}$. The horizontal and vertical axes correspond to the $\alpha$ and $\beta$ directions respectively (cf. Figure \ref{fig:test_projection}).}\label{fig:test_projection2}
\end{center}
\end{figure}

\subsection{Black Hole Shadows}
The appearance of the black hole shadow, which is the shadow cast by the photon capture surface of the black hole (and not the event horizon itself), can be determined by simply plotting all rays which are captured by the black hole. The black hole shadow profile is associated with the black hole spin and inclination angle of the observer, as studied in \cite[e.g.][]{fal00,tak04,cha13}. In Figure \ref{fig:shadow} we plot the black hole shadow for cases of a Schwarzschild ($a=0$) and a Kerr ($a=0.998$) black hole, when the observer is in the $x-y$ plane, i.e. $i=90^{\circ}$. For $a=0$, the shape of the black hole shadow, or more accurately the shadow of the photon capture region, follows the analytic description $\alpha^{2}+\beta^{2}=27$. The analytic solution for the general case, $a\neq0$, is discussed in e.g. \citet[][]{gre14,abd15}. The analytic solution (dashed red curves) are plotted in Figure \ref{fig:shadow} for comparison.

\begin{figure} 
\begin{center}
\includegraphics[scale=0.35,trim=1cm 0.2cm 1cm 0.5cm, clip=true,width=0.45\textwidth]{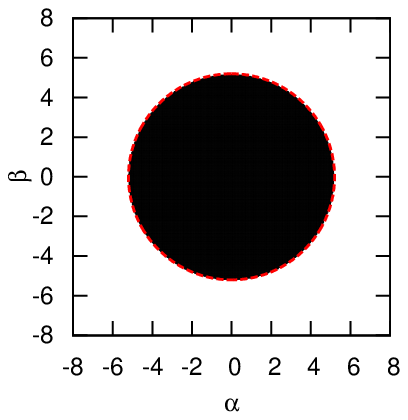}\\
\includegraphics[scale=0.35,trim=1cm 0.2cm 1cm 0.5cm, width=0.45\textwidth]{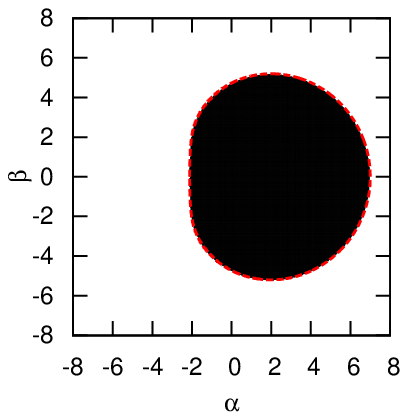}
\caption{Black hole shadow images for black holes with spin parameters of $a=0$ (top panel) and $a=0.998$ (bottom panel), as viewed along the equatorial plane. The dashed red curves indicate the analytic solution for the photon ring. The image resolution is $512\times512$ pixels.}\label{fig:shadow}
\end{center}
\end{figure}

\subsection{Keplerian Thin Disk}
We now consider an infinitesimally thin, Keplerian disk around a rotating black hole. The inner edge $r_{\rm in}$ of the disk is located at the marginally stable orbit \citep{bar72}, $r_{\rm ISCO}$ (innermost stable circular orbit). The outer edge, $r_{\rm out}$, is assumed to be located at $50 r_g$. For an observer inclination angle satisfying $\cos\theta_{\mathrm{obs}}=0.25$, the relative energy/frequency shift, $\gamma$, is shown in Figure \ref{fig:test_diskz} as a series of solid blue (blueshift) and solid red (redshift) contours. 
The approaching (left) side is blue shifted ($\gamma>1$) and the receding (right) side is red shifted ($\gamma<1$).
The dotted line denotes the region of zero redshift. The radial contours ($r$=constant, $cf.$ grid profile in bottom left panel of Figure \ref{fig:test_projection2}) are also shown as a series of concentric solid rings.

We next compute spectra from a multi-temperature disk as described in \citet{nov73} and \citet{pag74}.
The relation between the disk flux, $F$, and its effective temperature, $T_{\rm eff}$, may be written as:
\begin{equation}\label{eq:fr}
F(r)=\frac{3\dot{M}}{8\pi r^{2}}\frac{M}{r} f(r) = \sigma T_{\rm eff}^{4}(r) \ ,
\end{equation}
where $M$ is the mass of the black hole, $\dot{M}$ is the accretion rate, $f(r)$ is a correction factor related to the inner boundary of the disk and relativistic effects, and $\sigma$ is the Stefan-Boltzmann constant.
Under the assumption that the viscous stress vanishes at $r_{\rm ISCO}$,  the value of $f(r)$ (and hence $F(r)$)  also vanishes there.

The spectrum from a disk with $a=0.999$,  $i=85^{\circ}$, $M=10~M_{\odot}$, $\dot{M}=10^{19} \mathrm{g} \ \! \mathrm{s}^{-1}$, and (distance) $D = 10~\mathrm{kpc}$ is shown in  Figure \ref{fig:test_NTdisk}. The spectrum is plotted in terms of photon number flux density, $f_{\rm E}$ (photons/cm$^2$/s/keV), instead of energy flux density,  $F_{\nu}$ (ergs/cm$^2$/s/Hz)\footnote{$f_{\rm E}=1.51\times 10^{26} F_{\nu}/E$, where $E$ is in units of keV.}. 
We fix the disk inner radius at $r_{\rm in}=r_{\rm ISCO}$ and vary the outer radius of the disk as $r_{\rm out}=$ 50 $r_{g}$ and 500 $r_{g}$ to demonstrate how the turn over at lower energies depends on the disk's outer radius. Conversely, the high energy part of the spectrum manifests predominantly from regions with higher $T_{\rm eff}$ (i.e. from the inner edge of the disk) and therefore remains almost completely unchanged. 

Incorporated into the X-ray spectral fitting package \texttt{XSPEC}, the {\texttt{KERRBB}} model is a multi-temperature model of a geometrically-thin, steady accretion disk \citep[see][]{li05}. Disk self-irradiation, torque at the inner boundary of the disk and limb-darkening may be included within {\texttt{KERRBB}} by adjusting the appropriate model parameters.
For comparison, the result of the {\texttt{KERRBB}} model calculation with similar parameters is plotted as a dashed line in Figure \ref{fig:test_NTdisk} (for details, see the Figure caption).

\begin{figure} 
\begin{center}
\includegraphics[width=0.5\textwidth]{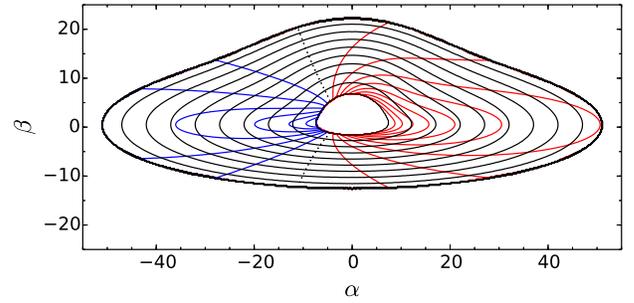}
\caption{Energy shift of a geometrically thin, Keplerian disk around a non-rotating black hole, with $\cos\theta_{\mathrm{obs}}=0.25$. The disk is rotating anticlockwise (i.e., in the $\phi$-direction). Only rays emitted directly from the disk are considered. The same contour values used in Figure 5.2 of \citet{ago97} are adopted: the radial contour plot shows $r/r_{\mathrm{g}}= 6.1, 11, 16,\cdots,41, 46, 50$ and the redshift contour plot shows $\gamma=0.55, 0.6,\ldots,0.95$ (red solid lines)$,1.0$ (black dotted line)$, 1.05,\ldots,1.25, 1.3$ (blue solid lines). This figure recovers Figure 5.2 of \citet{ago97}.}\label{fig:test_diskz} 
\end{center}
\end{figure}

\begin{figure} 
\begin{center}
\includegraphics[angle=-90,trim=0cm 0cm 0cm 0cm, clip=true,origin=c,width=0.5\textwidth]{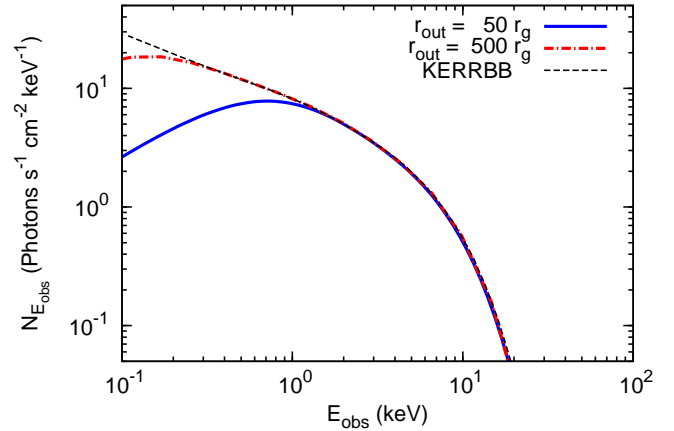} 
\caption{ Spectra from a relativistic multi-temperature disk described by \citet{nov73}, with $a=0.999$, $i=85^{\circ}$ and varying outer disk radius $r_{\rm out}$. The \texttt{KERRBB} model profile (dashed line) with similar parameters is obtained by using the \texttt{KERRBB} model in \texttt{XSPEC} with parameters [par1, par2,$\cdots$, par9]=[0, 0.999, 85, 10, 10, 10, 1, -1, -1], where the effects of torque at the inner edge of the disk, self-irradiation, and limb-darkening are omitted (see \href{http://heasarc.nasa.gov/xanadu/xspec/manual/XSmodelKerrbb.html}{http://heasarc.nasa.gov/xanadu/xspec/manual/XSmodelKerrbb.html} for definitions of the parameters).} \label{fig:test_NTdisk}
\end{center}
\end{figure}

\subsection{Keplerian Hot Spot}
A photon received by a distant observer at an observer time $t_{\rm obs}$ is emitted at a coordinate time, $t_{\rm emm}=t_{\rm obs}-\Delta t$, where $\Delta t$ is the time taken for the photon to travel from its point of emission to its point of reception on the observer image plane \citep[see][]{sch04}. 

The top panel of Figure \ref{fig:test_spectrogram} shows the spectrogram (time-dependent spectrum) of the direct emission from an orbiting hot spot rotating in a clockwise direction around the central black hole. The centre of the hot spot, $\mathbf{x}_{\rm spot}(t)$, orbits on the equatorial plane of a non-rotating black hole at $r_{\rm ISCO}$, observed at $i=60^{\circ}$. The emissivity is assumed to be a function of the distance ($d$) from the hot spot centre, where $d=\vert \mathbf{x}-\mathbf{x}_{\rm spot}(t)\vert$. Within $z>0$ and $d<4~r_{\rm spot}$, $r_{\rm spot}=0.5~r_{\mathrm{g}}$ and the emission is modelled as Gaussian in profile, in pseudo-Cartesian coordinates (equations (\ref{eq:x}), (\ref{eq:y}), and $z=r\cos\theta$)
\begin{equation}
 j(\mathbf{x}) \propto \exp{\left (-\frac{d^2}{2 r_{\rm spot}^2} \right)} \ .
\end{equation}
The hot spot is assumed to be optically thick, and therefore the computation is terminated once the ray intersects the hot spot surface. 
The shape and corresponding energy shift of the hot spot at specific orbital phases is shown in the bottom panel of Figure \ref{fig:test_spectrogram}.
The contribution of rays emitted from different locations of the hot spot at different local coordinate times, coupled with relativistic (e.g. length contraction and time dilation) and general-relativistic (e.g. gravitational lensing) effects, results in a distorted hot spot image \citep[see][]{you15}. Again, the emission region is magnified when the hot spot is `behind' the black hole, furthest away from the observer.
The lightcurve of the hot spot emission can be obtained by integrating the intensity over all frequency bins for each image, as shown in Figure \ref{fig:test_lightcurve}.
The spectrogram and lightcurve in Figure \ref{fig:test_spectrogram} recovers that found in Figures 4 and 6 of \citet[][]{sch04}.

\begin{figure} 
\begin{center}
\includegraphics[angle=-90,trim=0cm 2cm 1.2cm 2cm, clip=true,origin=c,width=0.5\textwidth]{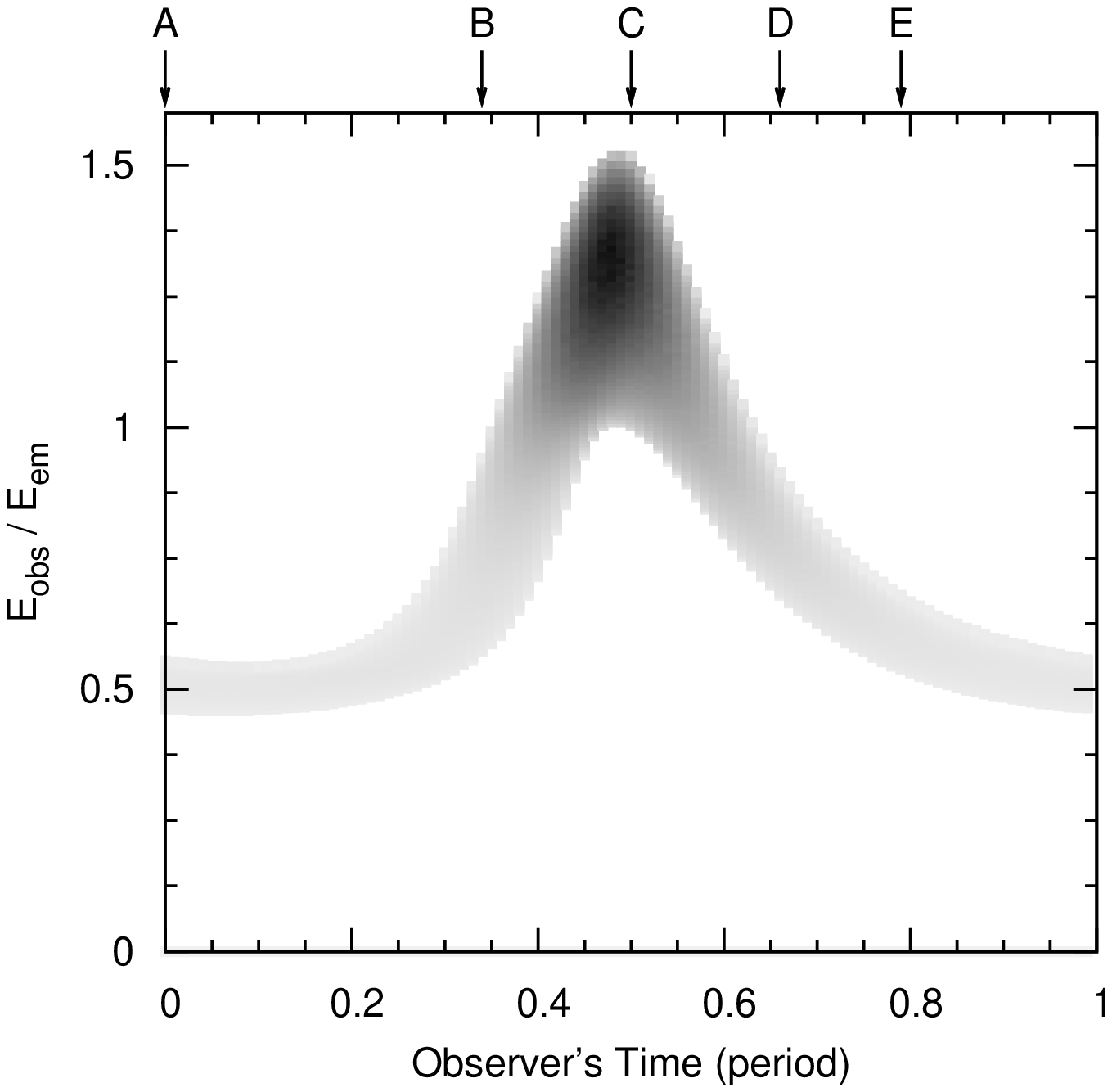} 
\includegraphics[width=0.45\textwidth]{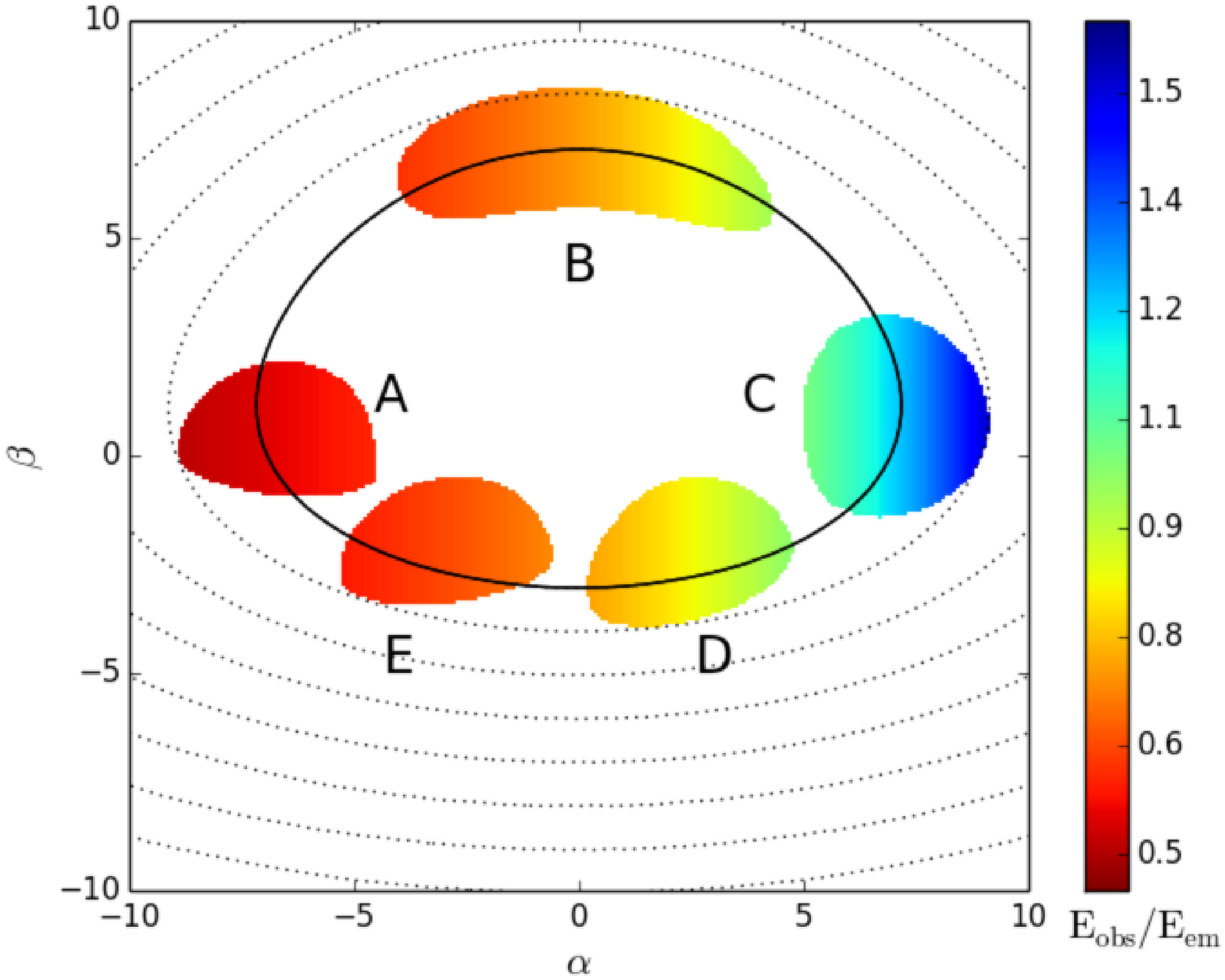}
\caption{Top panel: spectrogram of a circular hot spot with $r_{\rm spot}=0.5~r_{\mathrm{g}}$, orbiting a non-rotating black hole at $r_{\rm ISCO}$. The observer inclination angle is $i=60^{\circ}$. This result recovers Figure 4 of \citet{sch04}. Bottom panel: the shape and energy shift of the hot spot at specific observer times, coloured by energy shift, and also indicated by arrows A--E in the top panel.} \label{fig:test_spectrogram}
\end{center}
\end{figure}

\begin{figure} 
\begin{center}
\includegraphics[angle=-90,width=0.5\textwidth]{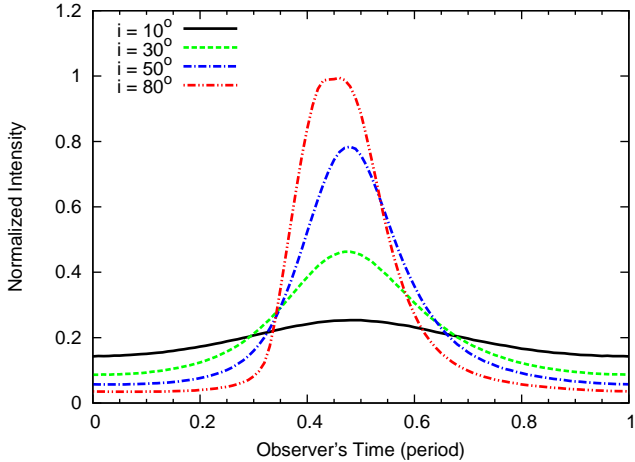}
\caption{Lightcurves of the hot spot described in Figure \ref{fig:test_spectrogram} for different observer inclination angles. The intensity is normalised to the sum of total intensity over one orbital period. This result is in agreement with Figure 6 of \citet{sch04}.  } \label{fig:test_lightcurve}
\end{center}
\end{figure}

\subsection{Keplerian Shell}

In order to test the radiative transfer formulation we now consider the emission from a Keplerian shell of plasma in rotation around a black hole \citep[see][]{bro06}. By setting $u^{\theta}=0$, the remaining components of the 4-velocity of the flow follow the same description as that of a Keplerian disk as given by, e.g., \citet[][]{cun75}. 
As a demonstrative calculation, we consider thermal synchrotron radiation from a distribution of relativistic electrons with an underlying relativistic Maxwellian profile. The angle-averaged emissivity coefficient for relativistic thermal synchrotron radiation may be written as \citep{mah96}:
\begin{equation}
j_\nu(T) = \nu \frac{4\pi ne^2}{\sqrt{3}c K_{2}(1/\Theta_\mathrm{e})}  M\left ( x_M\right ) \ , 
\end{equation}
where
\begin{equation}
M \left( x_M \right) = \frac{4.0505}{x_M^{1/6}} \left(1+\frac{0.40}{x_M^{1/4}}+\frac{0.5316}{x_M^{1/2}}\right)\exp(-1.8899x_M^{1/3}) \ .
\end{equation}
Here $x_M \equiv \nu / \nu_\mathrm{c}$ and
\begin{equation}
\nu_c = \left(\frac{3eB}{4 \pi m_e c}\right)\Theta_{e}^2 \ ,
\end{equation}
where $\Theta_e \equiv k_{\mathrm{B}}T/m_\mathrm{e} c^2$ is the dimensionless electron temperature and $K_\mathrm{n}$ is modified Bessel function of the second kind of order $\mathrm{n}$.
Since thermal synchrotron radiation is the only radiation source being considered, the absorption coefficient may be calculated via Kirchoff's law.

The Keplerian shell model has previously been used to simulate the image and spectrum of Sgr A$^{*}$ \citep[e.g.,][]{bro06,bro11}. We adopt the same self-similar electron number density profile and temperature profile given in equation (1) and Table 1 of \citet{bro06}, for the cases of black hole spin parameters of $a=0, 0.5$ and 0.998. The GRRT computation is performed within a shell of outer radius 500 $r_{\rm g}$, with a central black hole mass of $4\times10^{6}\;M_{\odot}$. The resulting images at 150 GHz, 340 GHz, and 1000 GHz are plotted in Figure \ref{fig:kep_imag}. The spectrum is shown in Figure \ref{fig:kep_spec}. It is clear that, besides the blackbody-like spectral component contributed by the thermal synchrotron emission, an additional power-law component is further needed to explain the observed data of Sgr A$^{*}$. On the other hand, because thermal synchrotron dominates the total emission of the image at the frequencies of interest, Figure \ref{fig:kep_imag} in general shows good agreement with Figure 1 of \citet{bro06}, in which the contribution from a non-thermal electron population is also included.

\begin{figure*} 
\begin{center}
\includegraphics[width=1.\textwidth]{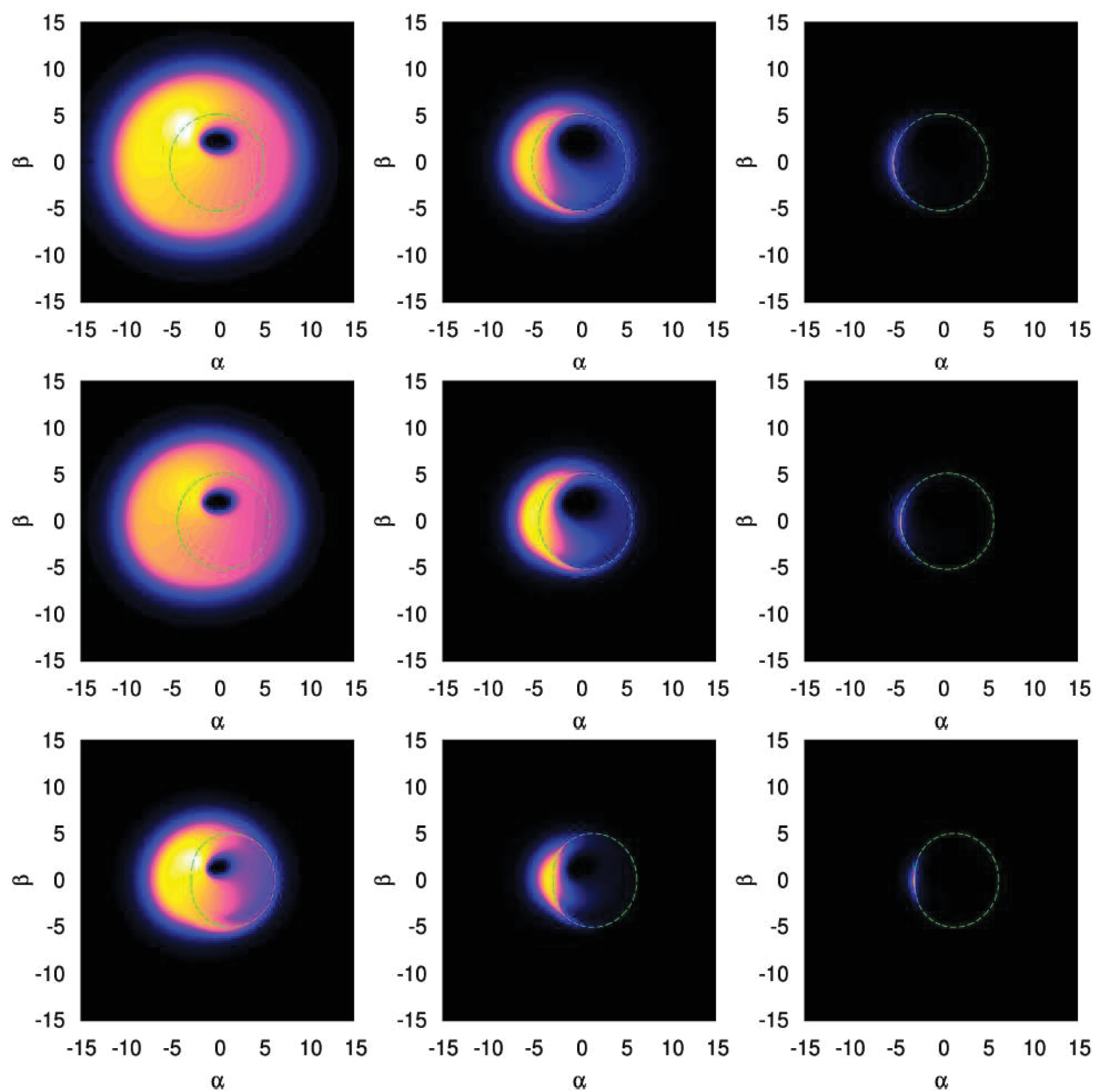}
\caption{Thermal synchrotron radiation image of a Keplerian shell around a black hole at 150 GHz (left column), 340 GHz (central column), and 1000 GHz (right column), as viewed at an observer inclination angle of $45^{\circ}$. From top row to bottom row, $a=0, 0.5$, and $0.998$ respectively.  The analytic solutions for the photon rings are shown by dashed green curves  for reference. The image intensity is plotted on a linear scale.} \label{fig:kep_imag}
\end{center}
\end{figure*}

\begin{figure}
\begin{center}
\includegraphics[scale=0.35,trim=0.2cm 0cm 0cm 0cm, clip=true,width=0.45\textwidth]{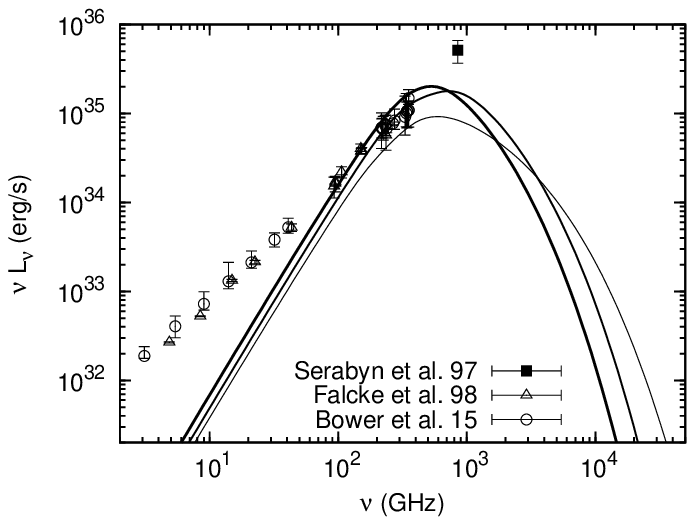}
\caption{Corresponding spectrum of the emission from a Keplerian shell around a non-rotating black hole as illustrated in Figure \ref{fig:kep_imag}. The cases $a=0, 0.5$, and $0.998$ are plotted as thick, medium and thin solid lines, respectively. Observational data from Sgr A$^{*}$ \citep[][]{ser97,fal98,bow15} are overlaid for comparison.} \label{fig:kep_spec}
\end{center}
\end{figure}

\section{Timing Benchmark}
To test the speed of \Odyssey we perform multiple runs on different image sizes for the same parameters as used to calculate the bottom  right panel of Figure \ref{fig:test_projection}, a benchmark also used in the GPU-based ray tracing code, \GRay \citep[][]{cha13}. 
The timing benchmark was performed a single NVIDIA Geforce GTX780 Ti graphics card (with 2880 CUDA cores and 3GB of GDDR5 RAM), and computed in double-precision floating-point arithmetic. In Table \ref{tab:timing}, we record the run time and total number of steps used for the Runge-Kutta method. Although the
run time and number of Runge-Kutta steps depends on the accuracy required by the adaptive step size \citep{pre92}, the
average time per Runge-Kutta step, per photon, remains roughly uniform with increasing numerical precision. 
Because the radiative transfer integration is also performed piecewise according to these steps, the average time per integration step, per photon provides a better standard for comparison with other ray-tracing codes with different underlying numerical schemes.
The value for the averaged run time per Runge-Kutta step, per photon as a function of image size is given in the last column of Table \ref{tab:timing}. 

We are also interested in how GPUs can boost the computation speed compared to conventional serial CPU and parallel CPU cluster codes with the same numerical algorithm.   
Therefore, we apply the same algorithm described in \S2 to a serial code and a parallel code in MPICH, which is a high-performance implementation of the Message Passing Interface (MPI) standard. These two codes are both written in C, using double-precision floating-point arithmetic. We compare the runtime result in Figure \ref{fig:speed}. The profile for the case of \Odyssey and MPICH flattens when integrating a small number of geodesics because the runtime is dominated by the time taken to invoke the CUDA or MPICH kernels. For small numbers of photons, the serial code outperforms the parallelised codes. However, parallel computations efficiently reduce the runtime for larger numbers of geodesics.  Comparing the unit price of a CPU cluster which can deliver computational results in an equal amount of time compared to one GPU card, it is clear that for the same unit price GPUs deliver results faster.

\GRay is a GPU-based ray-tracing code, based on the ray-tracing algorithm of \citet{psa12,bau12}. With the same benchmarks (lower right figure of Figure \ref{fig:test_projection}), \Odyssey and \GRay reveal similar profiles in runtime measures (comparing Figure \ref{fig:speed} with Figure 4 in \cite{cha13}). \Odyssey also reaches similar levels of performance to those reported by \GRay ($\sim$ 1 nanosecond per photon, per time step). The average runtime for \Odyssey can be less than one nanosecond when the image size is larger than $32^{2}$ pixels\footnote{The runtime varies from GPU to GPU according to the number of CUDA cores, as well as the number of GPUs.}. 

\begin{table}
\caption{Timing Performance of \Odyssey Code}
\label{tab:timing}
\begin{centering}
\begin{tabular}{cccc} 
\hline
\hline
Image size                & Run time$^{a}$  & Total R-K steps & Average time \\
(number of geodesics)  &   (ms)      &                        & (ns/step/photon)\\
\hline
$2^{2}$   & 56.921631  &5.718000e+003  &2488.703693\\
$4^{2}$  & 113.811523  &2.484900e+004  &286.257806\\
$8^{2}$  &116.409309  &9.876600e+004  &18.416211\\
$16^{2}$    &114.792221  &3.929330e+005  &1.141180\\
$32^{2}$   &117.429955  &1.548845e+006  &0.074041\\
$64^{2}$   &238.210617  &6.183181e+006  &0.009406\\
$128^{2}$   &605.967957  &2.468902e+007  &0.001498\\
$256^{2}$   &1873.401367   &9.866115e+007  &0.000290\\
$512^{2}$   &7090.669434  &3.944880e+008  &0.000069\\
$1024^{2}$   &25732.214844 & 1.577673e+009  &0.000016\\
\hline
\end{tabular} \\
\end{centering}
Notes. \\
$^{a}$ time required to compute the bottom  right panel of Figure \ref{fig:test_projection} by one nVIDIA GeForce GTX 780Ti graphics card with double-precision floating-point arithmetic. The GeForce GTX 780 Ti graphics card has 15 SMs and 192 cores per SM, giving 2880 CUDA cores in total.\\
\end{table}

\begin{figure} 
\begin{center} 
\includegraphics[width=0.45\textwidth]{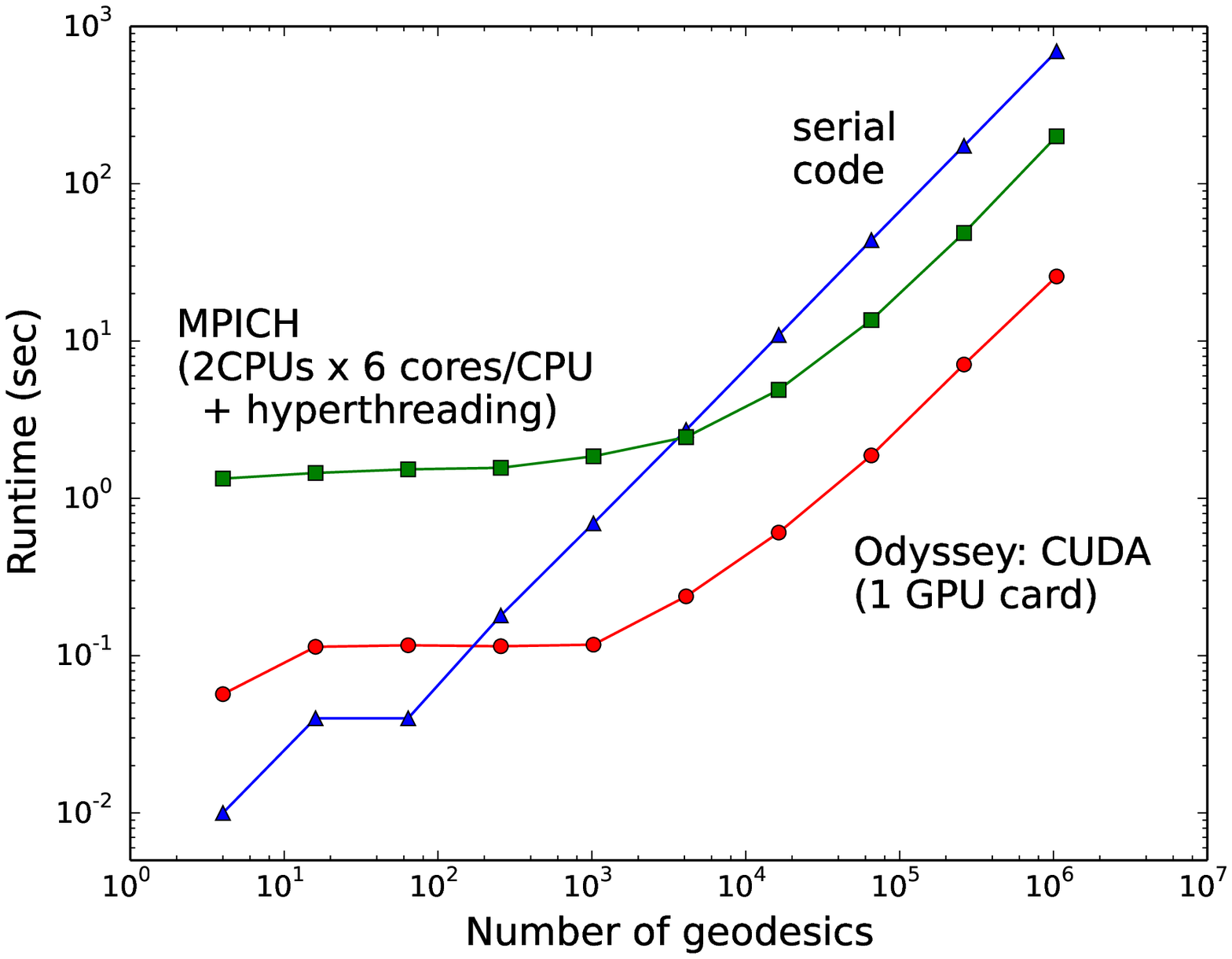}
\caption{Comparison of the computational time required to obtain the result in the bottom left panel of Figure \ref{fig:test_projection}, for an image size of $2^{n}\times2^{n}$ pixels ($n=1,2,\cdots,10$), using three different programming architectures: serial code (blue triangles), parallelised by MPICH (green squares), and CUDA (red circles). For small numbers of geodesics, the computational time is dominated by the time to launch the CUDA or MPICH kernel. The serial code is run on an Intel(R) Core(TM) i7-4940K 4.00GHz CPU. The MPICH code was run on  two Intel(R) Xeon(R) E5-2620 2.00GHz CPUs (each CPU has 6 cores), with hyperthreading enabled. Our GPU code (\Odyssey) was run on a single NVIDIA Geforce GTX 780Ti graphics card, as summarised in Table 1.}\label{fig:speed}
	\end{center}
	\end{figure}

\section{Summary and Discussion}

\begin{figure*} 
\begin{center}
\includegraphics[width=1.\textwidth]{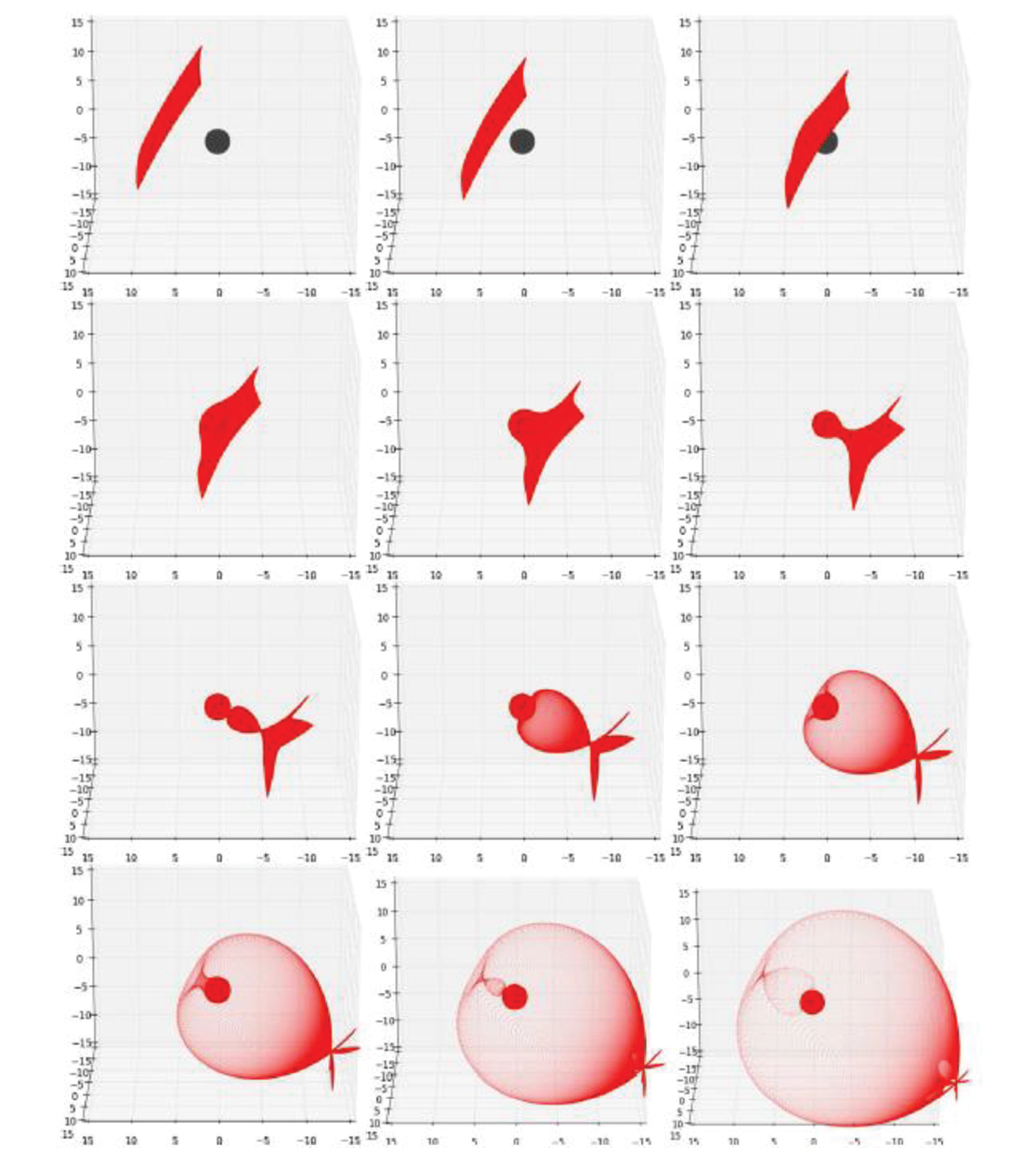}
\caption{Evolution of photon plane (backwards in time) near a non-rotating black hole at various coordinate times $t$, at which the emitted photons all reach the observer simultaneously. The time sequence is from left to right, top to bottom and the time interval between snapshots is $\mathrm{d}t=4$.} \label{fig:fpp}
\end{center}
\end{figure*}

\Odyssey is an accurate, flexible and efficient GPU-based code for ray-tracing and radiative transfer in the Kerr spacetime. Compared with \GRay \cite[][]{cha13}, \Odyssey adopts a different ray-tracing algorithm and CUDA code structure, with similar performance for the ray-tracing ($\lesssim$ 1 ns per step, per photon). The source code for \Odyssey is freely available at \href{https://github.com/hungyipu/Odyssey}{https://github.com/hungyipu/Odyssey}, 
with two default tasks for computing Figure 5 and the top middle panel of Figure \ref{fig:kep_imag}.
Users can easily modify the source code to suit their needs and efficiently perform GRRT calculations on their computer with a CUDA-capable GPU graphics card. 

An immediate and important application of \Odyssey is the rapid computation of images, spectra, and lightcurves of different black hole accretion and/or jet models, either provided semi-analytically \cite[e.g.,][]{bro06,bro09,pu15}, or numerically, e.g., from simulations of current state-of-the-art general-relativistic magnetohydrodynamic (GRMHD) simulation codes, such as HARM3D \citep[][]{nob09, mos14, mos15}, and RAISHIN \citep[][]{miz06}. 
Post-processing GRMHD simulation data for GRRT calculations has been considered in several studies to calculate the simulated spectrum and VLBI images from Sgr A* and M87 \citep[e.g.,][]{mos09,dex12,mos13}.
The works of \citet{cha15, cha15b} provide a good example of how GPUs can help to accelerate these computations significantly, one important application of which is to a time series of GRMHD data, in particular to extract the observed spectra and lightcurves from many different accretion models both accurately and efficiently. Such calculations are important for calculating the time-dependent emission from accretion onto black holes and warrant a detailed separate study.

Time-dependent GRRT involving a full consideration of the light-crossing time of each ray is needed when the light-crossing time-scale is comparable with the dynamical time-scale of the system, for example near the black hole event horizon or strong shock regions.
Starting from the observer's frame, the photon plane moving backward in time with the same increment $\Delta t$ resembles a plane on which all emitted photons will reach the observer simultaneously. We therefore term this plane the \FPP. At large distances, the \FPP resembles a conventional Euclidean plane because length-contraction and time-dilation effects are negligible far from the black hole. However, closer to the black hole, the \FPP become distorted, as demonstrated in Figure \ref{fig:fpp}. 
Although the GRRT integration along each ray between successive \textit{frozen photon planes} requires only the data within two planes,
near the black hole every photon on the same plane require different numbers of Runge-Kutta steps to reach the final photon plane (i.e. the distant observer). This is essentially the ``fast light" approximation, whereby it is assumed that all rays from everywhere within the computational time, at that particular observer time, arrive simultaneously. This enables ray-tracing on static time slices of GRMHD simulation data and is trivially parallelisable. 

Relaxing this approximation and considering the dynamical evolution of the medium as the ray propagates through it, along with the arrival time delays between neighbouring rays, is a significant computational challenge. Due to the limited amount of memory on-board each GPU, and the size of even modest 3D GRMHD data, post-processing GRMHD simulation results with time-dependent GRRT is not currently feasible with GPUs. However, recent progress in parallel computing, most notably with hybrid CPU and GPU programming architectures like \texttt{OpenCL} and \texttt{CUDA-Aware MPI}, offer several possibilities to approach this problem. Through sufficient load-balancing of the computation between CPUs and GPUs, one can in principle access the large amounts of RAM required for this task. We leave this to a future update of \Odyssey.

Finally, we note that \Odyssey is developed in the Microsoft Visual Studio environment, and as such it is possible to combine the \Odyssey algorithm with \texttt{DirectX} for visualising the propagation of rays in the Kerr spacetime. Together with the GUI, powered by \texttt{DirectX}, we also present a public software package and educational tool
 \OdysseyEdu, for demonstrating null geodesics around a Kerr black hole\footnote{available at \href{https://odysseyedu.wordpress.com/}{https://odysseyedu.wordpress.com/}}. 


\section*{acknowledgments}
\acknowledgments
We thank the anonymous referee for useful comments and suggestions which helped improve the manuscript.
We thank Steven V. Fuerst for sharing his code, Hsi-Yu Schive for helpful discussions about GPU programming, and K. Akiyama for discussions on spectral data from Sgr A$^{*}$. H.Y.P. and Z.Y. are grateful for numerous helpful and stimulating discussions with Kinwah Wu. H.Y.P. also thanks members of the GLT team for their support and encouragement. Z.Y.\, is supported by an Alexander von Humboldt Fellowship 
  and acknowledges support from the ERC Synergy Grant 
  ``BlackHoleCam -- Imaging the Event Horizon of Black Holes" (Grant 610058). 
S.J.Y. acknowledges support from the National Research Foundation of Korea to the Centre for Galaxy Evolution Research (No. 2010-0027910),
from the Mid-Career Researcher Program (No. 2015-008049) through the National Research Foundation (NRF) of Korea,
and from the Yonsei University Future-leading Research Initiative of 2014-2015.
This research has made use of NASA's Astrophysics Data System.


\end{document}